\documentclass[a4paper,12pt]{elsarticle}
\usepackage[utf8]{inputenc}
\usepackage{graphicx}
\title{Estimating performance of Feynman's ratchet with limited information}
\author{George Thomas\footnote{electronic address: georgethomas83@rediffmail.com}  
and Ramandeep S. Johal\footnote{electronic address: rsjohal@iisermohali.ac.in}}
\address{Department of Physical Sciences, \\ 
Indian Institute of Science Education and Research Mohali,\\
Sector 81, Knowledge City, Manauli P.O., Ajit Garh-140306, Punjab, India.}
\begin{document}
\begin{abstract}
We estimate the performance of Feynman's ratchet
at  given values of the ratio of cold to hot reservoir temperatures ($\theta$)
and the figure of merit (efficiency in the case of engine and  
coefficienct of performance in the case of refrigerator). The latter implies that only the ratio
of two intrinsic energy scales is known to the observer, but their exact values 
are completely uncertain. 
The prior probability distribution for the uncertain energy parameters is  
 argued to be Jeffreys' prior.
We define an average measure for performance of the model
by averaging, over the prior distribution, the power output (heat engine) or  
the $\chi$-criterion (refrigerator) which is the product of 
 rate of heat absorbed from the cold reservoir and the coefficient
of performance. 
We observe that the figure of merit, at optimal performance close to equilibrium,
is reproduced by the prior-averaging procedure. Further,  
we obtain the well-known expressions of finite-time thermodynamics
for the efficiency at optimal power and the coefficient
of performance at optimal $\chi$-criterion, given by
$1-\sqrt{\theta}$ and $1/\sqrt{1-\theta}-1$ respectively. 
This analogy is explored further and we point out
that the expected heat flow from and to the reservoirs,
behaves as an effective Newtonian flow. We also show, in
a class of quasi-static models of quantum heat engines,
how CA efficiency emerges in asymptotic limit with the use of
Jeffreys' prior. 
\end{abstract}
\maketitle
\section{Introduction} 
The benchmarks for optimal performance of heat engines and refrigerators, under reversible conditions, are
the carnot efficiency $\eta_c = 1-\theta$, and the carnot coefficient of performance $\zeta_c = \theta/ (1-\theta)$
respectively, where $\theta=T_2/T_1$ is the ratio of cold to hot temperatures of the reservoirs.
For finite-time models such as in the endoreversible approximation 
\cite{CA1975,DeVos1992,Salamon2001} and the symmetric
low-dissipation carnot engines \cite{Broeck2010}, the maximum power output 
is obtained at the so called Curzon-Ahlborn (CA) efficiency, $\eta^* = 1-\sqrt{\theta}$ \cite{CA1975}. 
However, CA-value is not as universal as $\eta_c$. For small temperature differences,
its lower order terms are obtained within the framework of linear
irreversible thermodynamics \cite{Broeck2005}.
Thus models with tight-coupling fluxes yield $\eta_c/2$  as the efficiency at maximum power.
Further, if we have a left-right symmetry, then the second-order 
term  $\eta_{c}^{2}/8$ is also universal \cite{Broeck2009}.

On the other hand, the problem of finding universal benchmarks for finite-time refrigerators 
is non-trivial. For instance, 
the rate of refrigeration ($\dot{Q}_2$), which seems a natural choice for optimization, cannot be optimized
under the assumption of a Newtonian heat flow ($\dot{Q} \propto \Delta T$)  between a reservoir 
and the working medium  \cite{Chen1990,Apertet}.
In that case, the maximum rate of refrigeration  is obtained 
as the coefficient of performance (COP) $\zeta$ vanishes. 
So instead, a useful target function $\zeta \dot{Q}_2$ has been 
used \cite{Chen1990, Armen2010, Tomas2012, Roco2012, Roco2013}, 
where $\dot{Q}_2$ is the heat absorbed per unit time by the working substance from the cold bath, 
or the rate of refrigeration.
The corresponding COP is found to be $\zeta^* = \sqrt{\zeta_c +1}-1$, for both the endoreversible
and the symmetric low-dissipation models. 
So this value is usually regarded as the analog of CA-value,
applicable to the case of refrigerators. 

In any case, the usual benchmarks for optimal performance  
of thermal machines are decided by recourse to optimization of
a chosen target function. The method also presumes a complete knowledge
of the intrinsic energy scales, so that, in principle, these scales can
be tuned to achieve the optimal performance. In this letter,
we present a different perspective on this problem. We
consider a situation where we have a limited or partial
information about the internal energy scales, so that we have
to perform an inference analysis \cite{Jeffreys1939} in order to estimate
the performance of the machine. Inference implies arriving at 
plausible conclusions assuming the truth of the given premises. 
Thus the objective of inference is not to predict the 
“true” behavior of a physical model but to arrive a rational
guess based on incomplete information.
In this context, the role
of prior information becomes central. In the spirit
of Bayesian probability theory, we treat all uncertainty
probabilistically and 
assign a prior probability distribution to the uncertain
parameters \cite{Jaynes1968}. We define an average or expected measure
of the performance, using the assigned prior distribution.
The approach was proposed
 by one of the authors \cite{Johal2010}
and has been then applied to  different models of  
heat engines \cite{GJ2012, GPJ2012, PJ2013, RJ2014}. 
These works show that CA-efficiency can be reproduced
as a limiting value when the prior-averaged work or power in a heat cycle
is optimized. In particular, for the problem of maximum work extraction
from finite source and sink, the behavior of efficiency at maximum
estimate of work shows universal features near equilibrium \cite{PJ2013}, 
e.g. $\eta = \eta_c/2 + \eta_{c}^2/8 + O[\eta_c^3]$.
 Similarly, other expressions for efficiency at maximum power, 
such as in irreversible models of stochastic engines 
\cite{Seifert, Zhang2006, BaratoSeifert}, 
which obey a different universality near equilibrium, can also be  
reproduced from the inference based approach \cite{GJ2012, JRM2014}.

However, so far the approach  has not been applied to other kinds of thermal machines such as
refrigerators. It is not obvious, beforehand, that the probabilistic approach 
can be useful in case of refrigerators also. 
The purpose of this paper is to extend the prior probability approach
by taking the paradigmatic Feynman's ratchet and pawl model \cite{Feynman1966}.
We show that the prior information infers not only the CA-efficiency $\eta^*$ in the engine mode,
but also the $\zeta^*$ value in the refrigerator mode of the model.
Further, we point out that the expected heat flows in the averaged model
behave as Newtonian flows. 

The present paper is organized as follows.
In Section 2, we describe the model of Feynman's ratchet as heat engine
and discuss its optimal configuration. In Section 2.1, the approach
based on prior information is applied to the case when the efficiency
of the engine is fixed, but the internal energy scales are 
uncertain. The approach is extended to the refrigerator mode,
in Section 3. In Section 4, we discuss alternate models where also
the use of Jeffreys' prior leads to emergence of CA efficiency.
Finally, Section 5 is devoted to discussion of 
results and conclusions.
\section{Optimal performance as a heat engine}
\begin{figure}[ht]
 \begin{center}
\includegraphics[width=8cm]{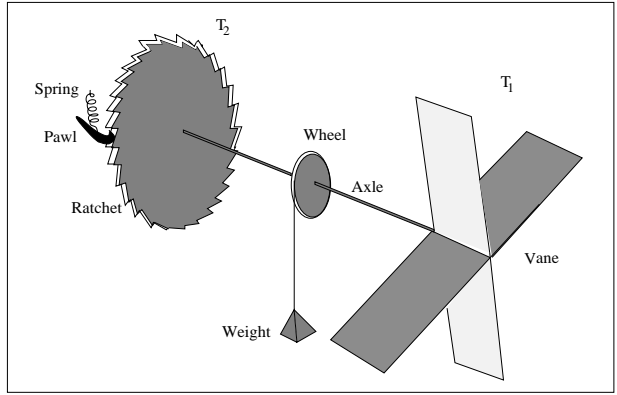}
\end{center}
\caption{A schematic of Feynman's ratchet.}
\label{fig-ratchet}
\end{figure}

The model of Feynman's ratchet as a heat engine consists of two heat
baths with temperatures $T_1$ and $T_2 (< T_1)$. 
A vane, immersed in the hot bath,  
is connected through an axle with a ratchet in contact with the cold bath, 
see Fig.\ref{fig-ratchet}.
The rotation of the ratchet is restricted in one direction due to 
a pawl which in turn is connected to a spring.
 The  axle  passes through the center of
a wheel from which hangs  a weight. So the directed 
motion of the ratchet rotates the wheel, thereby lifting the
weight. To raise the pawl,
 the system needs $\epsilon_2$ amount of energy
to overcome the elastic energy of the spring. Suppose that in each step,
the wheel rotate an angle $\delta$ and the torque
induced by the weight be $Z$. Then the system requires a 
minimum of $\epsilon_1=\epsilon_2+Z\delta$ energy to lift the weight.
Hence the rate of forward jumps for lifting  the weight is given as
\begin{equation}
 R_{\rm F}=r_0e^{-\epsilon_1/T_1},
\end{equation}
where $r_0$ is a rate constant and we have set Boltzmann's constant $k_{\rm B}=1$.

The statistical fluctuations can produce a directed motion at a finite rate, 
only if the ratchet-pawl system is mesoscopic. Hence the pawl
can undergo a Brownian motion by bouncing up and down as it is 
immersed in a finite temperature bath. 
This turns the wheel in
 backward direction and lowers the position of the 
weight. This is the reason that the system cannot work as an engine if $T_1=T_2$
\cite{Feynman1966}.

The rate of the backward jumps is
\begin{equation}
 R_{\rm B}=r_0e^{-\epsilon_2/T_2}.
\end{equation}
Thus one can regard $Z\delta$ and $-Z\delta$  as the work done by and on the system, respectively.
In an infinitesimally small time interval $\delta t$, the work done by
the system is given as
\begin{eqnarray}
 W&=&(\epsilon_1-\epsilon_2) (R_{\rm F}-R_{\rm B}) \delta t,\nonumber \\
 &=&r_0(\epsilon_1-\epsilon_2) \left(e^{-\epsilon_1/T_1}- e^{-\epsilon_2/T_2}\right)\delta t.
 \label{work-ratchet}
\end{eqnarray}
Thus the power output of the engine is defined as $P = W/\delta t$.
Similarly, the rate of heat absorbed from the hot reservoir, is  given as
\begin{equation}
\dot{Q}_1=r_0\epsilon_1 \left(e^{-\epsilon_1/T_1}- e^{-\epsilon_2/T_2}\right),
\label{q1dot}
\end{equation}
or the amount of heat absorbed in the small time interval is $Q_1 = \dot{Q}_1 \delta t$.
Then the efficiency of the engine is given by 
\begin{equation}
 \eta=\frac{W}{Q_1}=1-\frac{\epsilon_2}{\epsilon_1}.
 \label{efficiency-ratchet}
\end{equation}
The rate at which waste heat is rejected to the cold reservoir is $\dot{Q}_2 = \dot{Q}_1 -P$,
which follows from the conservation of energy flux.

The power output, optimized with respect to energy scales $\epsilon_1$ and $\epsilon_2$  
\cite{Hernandez, Tu2008}, is given by
\begin{equation}
 \tilde{P}=r_0 e^{-1} T_1 \eta_c^2 (1-\eta_c)^{(1-\eta_c)/\eta_c}.
 \label{MaxPower}
\end{equation}
The corresponding efficiency at maximum power is
\begin{equation}
\tilde{\eta}=\frac{\eta_c^2}{\eta_c-(1-\eta_c)\ln{(1-\eta_c)}}. 
\label{efopt}
\end{equation}
Further, it was discussed in Ref. \cite{Tu2008} that the above expression
for efficiency shares some universal properties of efficiency
at optimal power found in other finite-time models \cite{CA1975,Seifert}. 
\subsection{Prior information approach}
Now we consider a situation where the efficiency of the engine has some pre-specified value $\eta$,
but the energy scales ($\epsilon_1,\epsilon_2$)
are not given to us in a priori information. Since $\eta$ is known, the problem is reduced to
a single uncertain parameter, due to Eq. (\ref{efficiency-ratchet}).
One can cast the problem either in terms of $\epsilon_1$ or $\epsilon_2$.
In terms of the latter, we can write power as
\begin{equation}
P(\eta,\epsilon_2)= \frac{r_0 \eta \epsilon_2}{(1-\eta)}
\left(e^{-\epsilon_2/(1-\eta)T_1}- e^{-\epsilon_2/T_2}\right).
\end{equation}
Analogous to quantification of prior information in  Bayesian statistics, 
we assign a prior probability distribution for 
$\epsilon_2$ in some arbitrary, but a finite range of positive values:
$[\epsilon_{\rm min},\epsilon_{\rm max}]$.  Later we consider an asymptotic
range in which the analysis becomes simplified and 
we observe universal features.

Now consider two observers $A$ and $B$ who respectively assign a  prior for $\epsilon_1$  and $\epsilon_2$.
Taking the simplifying assumption that each observer is in an equivalent state of knowledge,
we can write \cite{Jaynes1968,PJ2013}
\begin{equation}
 \Pi(\epsilon_1)=\Pi(\epsilon_2)\left|\frac{d\epsilon_2}{d\epsilon_1}\right|,
 \label{choice-of-prior}
\end{equation}
where $\Pi$ is the prior distribution function, taken to be of the same form for each observer.
At a fixed known value of efficiency, it implies that
$\Pi(\epsilon_2)= N/\epsilon_2$, where the normalization constant,
$N = \left[\ln \left({\epsilon_{\rm max} }/
{\epsilon_{\rm min}}\right) \right]^{-1}$.   This is also known as Jeffreys' prior 
for a one-dimensional scale parameter \cite{Jeffreys1939, Jaynes1968, Abe2014}.

Now the expected value of power, over this prior, is defined to be 
 \begin{eqnarray}
\overline{P}(\eta)&=& \int_{\epsilon_{\rm min}}^{\epsilon_{\rm max}}
P(\eta,\epsilon_2) \Pi(\epsilon_2)d\epsilon_2\nonumber \\
&=&\frac{C \eta}{(1-\eta)}\int_{\epsilon_{\rm min}}^{\epsilon_{\rm max}}
\left(e^{-\epsilon_2/(1-\eta)T_1}- e^{-\epsilon_2/T_2}\right)d\epsilon_2,
\end{eqnarray}
where 
\begin{equation}
C=r_0\left[\ln \left(\frac{\epsilon_{\rm max} }{\epsilon_{\rm min}}\right) \right]^{-1}. 
\label{defc}
\end{equation}
Upon performing the integration, we get
 \begin{eqnarray}
\overline{P}(\eta)&=&CT_1\eta\left(e^{-\epsilon_{\rm min}/(1-\eta)T_1}-
e^{-\epsilon_{\rm max}/(1-\eta)T_1}\right)\nonumber\\
&&+\frac{CT_2 \eta}{(1-\eta)}\left(e^{-\epsilon_{\rm max}/T_2}
- e^{-\epsilon_{\rm min}/T_2}\right).
\end{eqnarray}
Now this expected power depends on the extreme values 
defining the range of the prior. We chose a finite range in order to
define a normalized prior distribution. Otherwise, information on the finite values of 
these scales  is not available. 
On the other hand, as the range is made arbitrarily large, the average power becomes
increasingly small. Thus a comparison between the absolute magnitudes of optimal power
(Eq. (\ref{MaxPower}))
and the prior-averaged power does not seem fruitful. However,
the expected power is seen to become optimal at a certain value of the given efficiency.
Further, universal features are shown by this efficiency in the asymptotic
limit. It also provides a good estimate of the actual values of efficiency at
maximum power. 

\begin{figure}[ht]
 \begin{center}
\includegraphics[width=7.5cm]{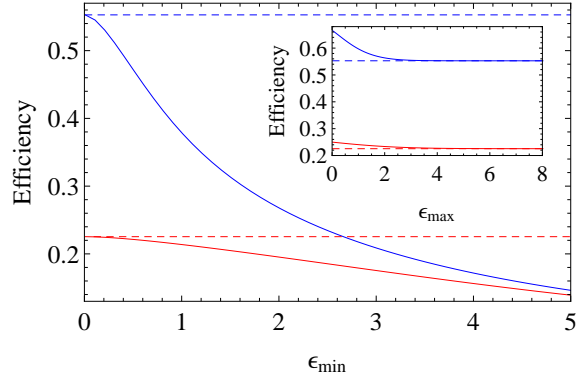}
\end{center}
\caption{The efficiency at maximum expected power is plotted versus $\epsilon_{\rm min}$ (scaled by $T_1$),
while $\epsilon_{\rm max}=10$.
The upper and lower curves correspond to $\theta=0.2$ and $\theta=0.6$, respectively.
The dashed lines represent corresponding CA values. 
The efficiency is also plotted versus $\epsilon_{\rm max}$ (inset),  assuming 
$\epsilon_{\rm min}=0.01$. 
For larger values of $\epsilon_{\rm max}$, the efficiency approaches CA value.
}
\label{finite-limits-engine}
\end{figure}

Hence, on maximizing 
$\overline{P}(\eta)$ with respect to $\eta$, we get
 \begin{eqnarray}
\frac{\partial \overline{P}}{\partial \eta}&\equiv&  T_1\left(e^{-\epsilon_{\rm min}/(1-\eta)T_1}
- e^{-\epsilon_{\rm max}/(1-\eta)T_1}\right)\nonumber\\
&&- \frac{\eta}{(1-\eta)^2}\left(\epsilon_{\rm min}e^{-\epsilon_{\rm min}/(1-\eta)T_1} 
- \epsilon_{\rm max}e^{-\epsilon_{\rm max}/(1-\eta)T_1}\right)\nonumber\\
&&+ \frac{T_2}{(1-\eta)^2}\left(e^{-\epsilon_{\rm max}/T_2}- e^{-\epsilon_{\rm min}/T_2}\right)=0.
\label{deta}
\end{eqnarray}
For given values of the limits, we obtained numerical solution for $\eta$.
As shown in Fig. \ref{finite-limits-engine}, the  efficiency at maximum expected
power versus  $\epsilon_{\rm min}$ is plotted, 
for a given value of  
the upper limit $\epsilon_{\rm max}$.
Alternately, setting the lower limit
$\epsilon_{\rm min}$ as relatively small in magnitude, one can visualise 
the behaviour of the efficiency with $\epsilon_{\rm max}$. Interestingly, these 
solutions show convergence to the CA-value, $1-\sqrt{\theta}$. 
\par
The convergence to the CA value as observed in Fig. 1, can be argued as follows.
Let us assume that the temperature gradient is not very large, i.e. $\theta$
is not close to zero. Or in other words, $\eta_c$ is small compared to unity.
This implies that $\eta$ is also small since it is bounded from above by $\eta_c$.
%
Now let us consider the limits which satisfy, $\epsilon_{\rm max}>>T_1$
and $\epsilon_{\rm min} << T_2$ \cite{GJ2012},  referred to as
{\it asymptotic range} in the following. Then the condition (\ref{deta}) simplifies to the form
\begin{equation}
T_1-\frac{T_2}{(1-\eta)^2}=0.
\end{equation}
This implies that the efficiency at optimal $\overline{P}$, approaches the CA value. 

\begin{figure}[ht]
 \begin{center}
\includegraphics[width=8cm]{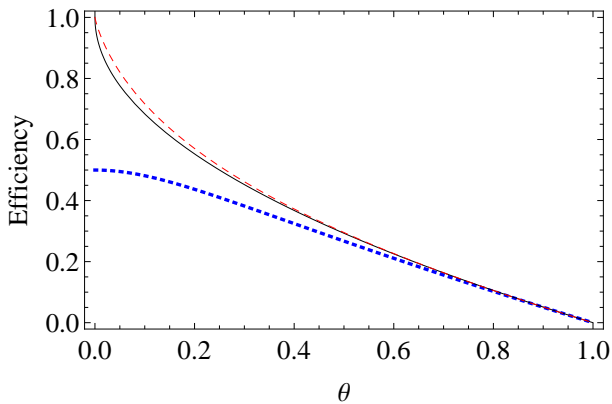}
\end{center}
\caption{The solid curve shows the CA value obtained for Feynman's ratchet
at optimal expected power, using Jeffreys' prior in the asymptotic range. 
The dotted curve is the corresponding efficiency 
(Eq. (\ref{unieff})) when a uniform prior is used, also in the asymptotic range.
 The dashed curve represents the efficiency at optimal power $\tilde{\eta}$,
 Eq. (\ref{efopt}).}
\label{fig-ratchet-engine}
\end{figure}

{\it Uniform Prior}: On the other hand, maximal ignorance about the likely values of a parameter may be
represented by a uniform prior density,
$\Pi_u=1/(\epsilon_{\rm max}-\epsilon_{\rm min})$.
Then the  expected power, is given as 
 \begin{eqnarray}
\overline{P}_u(\eta)&=& \int_{\epsilon_{\rm min}}^{\epsilon_{\rm max}}
P(\eta,\epsilon_2) \Pi_u(\epsilon_2)d\epsilon_2\nonumber \\
&=&\frac{C' \eta}{(1-\eta)}\int_{\epsilon_{\rm min}}^{\epsilon_{\rm max}}
\epsilon_2\left(e^{-\epsilon_2/(1-\eta)T_1}- e^{-\epsilon_2/T_2}\right)d\epsilon_2,
\end{eqnarray}
where $C'=r_0/(\epsilon_{\rm max}-\epsilon_{\rm min})$. Integrating the above equation, we get 
\begin{eqnarray}
 \overline{P}_u(\eta)&=&C'T_1\eta \left[ (T_1(1-\eta)+\epsilon_{\rm min})
 e^{-\epsilon_{\rm min}/(1-\eta)T_1} \right.\nonumber\\
 &&\left.-(T_1(1-\eta)+\epsilon_{\rm max})e^{-\epsilon_{\rm max}/(1-\eta)T_1}\right.\nonumber\\
 &&\left.-\frac{T_2}{(1-\eta)T_1}\left((T_2+\epsilon_{\rm min})e^{-\epsilon_{\rm min}/T_2}
 -(T_2+\epsilon_{\rm max})e^{-\epsilon_{\rm max}/T_2} \right) \right]. 
\end{eqnarray}
Here, we are interested in the efficiency at maximum expected power ($\eta_u$)
in the asymptotic range. Therefore, by putting
$\partial\overline{P}_u(\eta)/\partial \eta=0$ and then considering the asymptotic limit, we get
\begin{equation}
(1-\eta_u)^2(1-2 \eta_u)T_1^2  - T_2^2=0,
\end{equation}
whose real solution is given by
\begin{equation}
 \eta_u=\frac{1}{6K}(5K-K^2-1),
 \label{unieff}
\end{equation}
where  $K =(1+54\theta^2+6\sqrt{3}\theta \sqrt{1+27\theta^2})^{1/3}$.
These efficiencies are compared in Fig. \ref{fig-ratchet-engine}.
In particular, we note that in the asymptotic range, the efficiency depends only
on the ratio of the reservoir temperatures.
Further, the use of Jeffreys' prior gives a closer approximation 
to the actual behavior of efficiency at optimal performance
of the engine.
\par
To compare these efficiencies near equilibrium i.e. $\eta_c$ close to zero,  we 
expand these expressions as Taylor series for small values of $\eta_c$,
\begin{eqnarray}
\tilde{\eta}&=&\frac{\eta_c}{2}+\frac{\eta_c^2}{8}+\frac{7 \eta_c^3}{96}+O[\eta_c^4]
\;\; ({\rm Eq. (\ref{efopt})}; \mbox{ at optimal power}) \label{seropt}\\
\eta^*&=&\frac{\eta_c}{2}+\frac{\eta_c^2}{8}+\frac{6 \eta_c^3}{96}+O[\eta_c^4]\; \; {\rm (CA\; value\; from\; 1/\epsilon_2 \; prior)} \label{serp} \\
\eta_u&=&\frac{\eta_c}{2}+\frac{\eta_c^2}{16}+\frac{\eta_c^3}{64}+O[\eta_c^4].\;\; {\rm (with\; uniform\; prior)}
\end{eqnarray}
The series in Eqs. (\ref{seropt}) and  (\ref{serp}) were obtained in Ref. \cite{Tu2008}.
We note that $\eta_c/2$ term in the optimal performance can be faithfully reproduced by the expected power 
irrespective of the chosen prior.
However, the second order term follows from the use of Jeffreys' prior.   
\section{Optimal performance as a refrigerator}
In this section, we consider the function of 
Feynman's ratchet as a refrigerator \cite{He2013, Tu2014, Liu2006, Chen2009}. It is analogous to  
Büttiker-Landauer model \cite{Buttiker1987, Landauer1988}, as discussed in \cite{Tu2014}.
By optimizing the target function $\chi = \zeta \dot{Q}_2$  for Feynman's ratchet, the COP at optimal
performance $\tilde{\zeta}$ satisfies a transcendental equation \cite{Tu2014}.
The solution can be approximated by an interpolation formula 
\begin{equation}
\tilde{\zeta}=\sqrt{\zeta_c+(0.954)^2}-0.954.
\end{equation}
Similar to the case of heat engine,  
we now show using the prior based approach, 
that COP at optimal performance can be obtained
for Feynman's ratchet as refrigerator. The COP for certain values of $\epsilon_1$ and $\epsilon_2$
is given by $\zeta = \epsilon_2/ (\epsilon_1- \epsilon_2)$.
Also the rate of refrigeration is given by
\begin{equation}
\dot{Q}_2= r_0 \epsilon_2 \left(e^{-\epsilon_2/T_2}- e^{-\epsilon_1/T_1}\right). 
\end{equation}
In terms of $\zeta$ and one of the scales say, $\epsilon_2$, the $\chi$-criterion is
given by
\begin{equation}
\chi(\zeta,\epsilon_2)=\zeta r_0 \epsilon_2 \left(e^{-\epsilon_2/T_2}- 
e^{-\epsilon_2(1+\zeta)/\zeta T_1}\right).
\end{equation}
Now we suppose that the COP is fixed at some value $\zeta$, and 
$\epsilon_2$ is uncertain, within the range $[\epsilon_{\rm min},\epsilon_{\rm max}]$.
Then Jeffreys' prior for $\epsilon_2$ can be argued, similar to Eq. (\ref{choice-of-prior}).
Now we define the expected value of $\chi$ as
\begin{eqnarray}
 \overline{\chi}(\zeta)&=&\int_{\epsilon_{\rm min}}^{\epsilon_{\rm max}}\chi(\zeta,\epsilon_2)\Pi(\epsilon_2)d\epsilon_2\\
 &=&C
 \int_{\epsilon_{\rm min}}^{\epsilon_{\rm max}}
 \zeta\left(e^{-\epsilon_2/T_2}- e^{-\epsilon_2(1+\zeta)/\zeta T_1}\right)d\epsilon_2,
\end{eqnarray}
where $C$ is given by Eq. (\ref{defc}). Upon integrating the above equation, we get
\begin{eqnarray}
 \overline{\chi}(\zeta)&=&C\zeta T_2\left(e^{-\epsilon_{\rm min}/T_2}- e^{-\epsilon_{\rm max}/T_2}\right)\nonumber\\
 &&+ \frac{C\zeta^2T_1}{(1+\zeta)}
 \left(e^{-\epsilon_{\rm max}(1+\zeta)/\zeta T_1}- e^{-\epsilon_{\rm min}(1+\zeta)/\zeta T_1}\right).
\end{eqnarray}
 As with power output for the engine, the average $\overline{\chi}$
becomes increasingly small in the asymptotic limit.
In the following, we focus on COP at maximal $\overline{\chi}$, in the 
asymptotic limit.

\begin{figure}[ht]
 \begin{center}
\includegraphics[width=8cm]{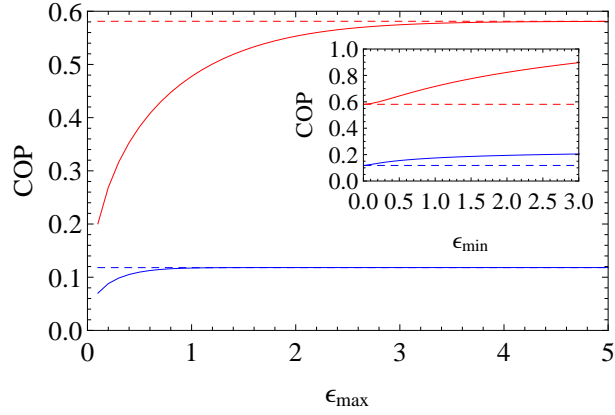}
\end{center}
\caption{The COP at maximum expected $\chi$-criterion is plotted versus $\epsilon_{\rm max}$ (scaled by $T_1$),
while $\epsilon_{\rm min}=0.01$.
The upper and lower curves correspond to $\theta=0.6$ and $\theta=0.2$, respectively. The dashed lines
represent corresponding $\zeta^*$ values. For larger values of $\epsilon_{\rm max}$, COP approaches
the corresponding $\zeta^*$. In inset, COP is plotted versus $\epsilon_{\rm min}$,  when
$\epsilon_{\rm max}=10$. The COP approaches $\zeta^*$ as $\epsilon_{\rm min}$ takes smaller values.
}
\label{finite-limits-refri}
\end{figure}

\begin{figure}[ht]
 \begin{center}
\includegraphics[width=8cm]{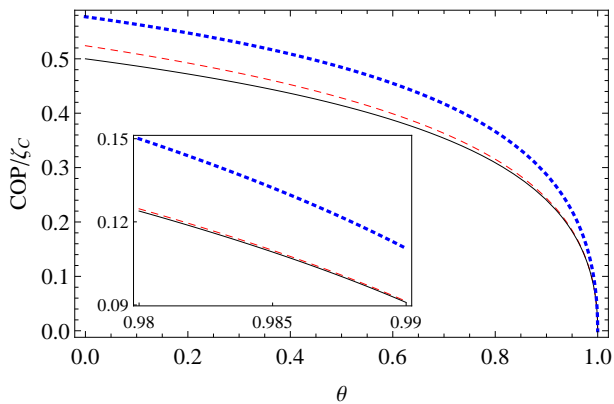}
\end{center}
\caption{The COP (scaled by the Carnot value $\zeta_c$) is plotted versus $\theta$.
The solid curve shows the COP at  optimal expected performance ($\overline{\chi}$) 
when Jeffreys' prior is assigned and the asymptotic range is applied.
The dashed line represents the interpolation formula for COP corresponding to
the optimum  $\chi$ value \cite{Tu2014}. The top, dotted line
is the result of uniform prior, again in the asymptotic range.
The inset shows the same three quantities
for close-to-equilibrium values of $\theta$.}
\label{fig-fridge}
\end{figure}

So the maximum of $\overline{\chi}$ with respect to $\zeta$, is evaluated as 
\begin{eqnarray}
 \frac{\partial\overline{\chi}}{\partial\zeta}&\equiv&T_2
 \left(e^{-\epsilon_{\rm min}/T_2}- e^{-\epsilon_{\rm max}/T_2}\right)\nonumber\\
 &&+ \frac{\zeta(\zeta+2)T_1}{(1+\zeta)^2}
 \left(e^{-\epsilon_{\rm max}(1+\zeta)/\zeta T_1}- 
 e^{-\epsilon_{\rm min}(1+\zeta)/\zeta T_1}\right)\nonumber\\
  && + \frac{1}{(1+\zeta)}
 \left(\epsilon_{\rm max} e^{-\epsilon_{\rm max}(1+\zeta)/\zeta T_1}-\epsilon_{\rm min}
 e^{-\epsilon_{\rm min}(1+\zeta)/\zeta T_1}\right)=0.
\end{eqnarray}
The numerical solution for $\zeta$ versus one of the limits is shown in Fig. \ref{finite-limits-refri}. 
Finally, in the asymptotic range, the above expression reduces to
\begin{equation}
\frac{\zeta(\zeta+2)}{(1+\zeta)^2}-\frac{T_2}{T_1}=0.
\end{equation}
So the permissible solution ($\zeta >0$)  of the above quadratic equation,
which maximizes $\overline{\chi}$, is given as
 \begin{eqnarray}
  \zeta^* &=&\frac{1}{\sqrt{(1-\theta)}}-1, \nonumber \\
          &=& \sqrt{1 + \zeta_c}  -1.
 \end{eqnarray}
{\it Uniform Prior}:
On the other hand, with uniform prior, the expected $\chi$-criterion is given as
\begin{eqnarray}
 \overline{\chi}_u(\zeta)&=&\int_{\epsilon_{\rm min}}^{\epsilon_{\rm max}}
 \chi(\zeta,\epsilon_2)\Pi_u(\epsilon_2)d\epsilon_2\nonumber\\
 &=&C'\zeta
 \int_{\epsilon_{\rm min}}^{\epsilon_{\rm max}}
 \epsilon_2\left(e^{-\epsilon_2/T_2}- e^{-\epsilon_2(1+\zeta)/\zeta T_1}\right)d\epsilon_2.
\end{eqnarray}
Upon integrating  the above equation,  we get
\begin{eqnarray}
 \overline{\chi}_u(\zeta)&=& C'\zeta \left[ T_2\left((T_2+\epsilon_{\rm min})e^{-\epsilon_{\rm min}/T_2}
 -(T_2+\epsilon_{\rm max})e^{-\epsilon_{\rm max}/T_2} \right) \right.\nonumber\\
 && \left.+\frac{T_1}{(1+\zeta)^2}(\epsilon_{\rm max}(1+\zeta)+\zeta T_1)e^{-\epsilon_{\rm max}(1+\zeta)/\zeta T_1}\right.\nonumber\\
 &&\left.-\frac{T_1}{(1+\zeta)^2}(\epsilon_{\rm min}(1+\zeta)+\zeta T_1)e^{-\epsilon_{\rm min}(1+\zeta)/\zeta T_1}\right].
\end{eqnarray}
Now, we want to  estimate $\zeta_u$, the COP 
at maximum expected $\chi$-criterion in asymptotic range.
Hence, by putting $\partial\overline{\chi}_u(\zeta)/\partial \zeta=0$ 
and imposing the asymptotic range, we obtain the following equation
\begin{equation}
 T_2^2(1+\zeta_u)^3-T_1^2\zeta_u^2(3+\zeta_u)=0,
\end{equation}
whose acceptable solution can be finally written in the following form
 \begin{equation}
  \zeta_u = \frac{2}{\sqrt{1-\theta^2}} \cos\left[ \frac{\pi}{3} - 
  \frac{\sin^{-1} \theta}{3}   \right]-1.
\label{unicop}
 \end{equation}
 Again, we see that in the asymptotic range, the COP is given only in terms
 of the ratio of the reservoir temperatures.
We show in Fig. \ref{fig-fridge},  a comparison amongst the different expressions
 for COP at optimized performance versus this ratio. 
\par
In near-equilibrium regime, the Carnot COP $\zeta_c$, as well as $\zeta^*$ become large in magnitude.
One can then write the series expansion for $\zeta^*$ relative to $\zeta_c$ as follows:
\begin{equation}
\frac{\zeta^*}{\zeta_c} = \frac{1}{\sqrt{\zeta_c}} - \frac{1}{\zeta_c} + O[\zeta_{c}^{-3/2}].
 \end{equation}
In this case, $\zeta_u$ relative to $\zeta_c$ behaves as follows:
\begin{equation}
\frac{\zeta_u}{\zeta_c} = \sqrt{\frac{3}{2{\zeta_c}}} - \frac{4}{3 \zeta_c} + O[\zeta_{c}^{-3/2}].
 \end{equation}
According to Refs. \cite{Tu2014, ShengTu} close to equilibrium and upto the leading order, 
${\tilde{\zeta}}/{\zeta_c}$
behaves as $1/\sqrt{\zeta_c}$. The optimal behavior is thus reproduced by the use of
Jeffreys' prior, but uniform prior is not able to generate this dependence. 
Similarly, for large temperature differences,  $\zeta_c \to 0$, 
we get the limiting behavior as $\zeta^*/\zeta_c \to 1/2$ while $\zeta_u/\zeta_c \to 1/\sqrt{3}$.
The interpolation formula at optimal performance, gives
$\tilde{\zeta}/ \zeta_c \to 0.524$  \cite{Tu2014}.

Before closing this section, we point out that performing the same analysis in terms of $\epsilon_1$
as the uncertain scale, we obtain a similar behavior 
in the asymptotic range of values, and the same figures of merit, $\eta^*$ and $\zeta^*$,
are obtained with the choice of Jeffreys' prior.
\section{Other Models}
So far, we have focused on the performance of Feynman's ratchet.
In the following, we wish to point out that the above inference
analysis can also be performed on other
classes of heat engines/refrigerators \cite{Johal2010}. The model which we discuss
below is a four-step heat cycle performed by a few-level quantum
system (working medium). Further, the cycle is accomplished
using infinitely slow processes. The particular cycle is
the quantum Otto cycle \cite{Kieu2004, Quan2007}. 

Consider a quantum system with Hamiltonian $H_1
=\sum_{n=1}^{M} \varepsilon_{n}^{(1)}
\vert n\rangle \langle n\vert$, with 
eigenvalue spectrum of the form  $\varepsilon_{n}^{(1)}= \varepsilon_n a_1$. Here 
$\varepsilon_n$ is characterised by the energy quantum number 
and other parameters/constants which remain fixed during the cycle.
We assume there are $M$ non-degenerate levels. The parameter 
$a_1$ represents an external control, equivalent to 
applied magnetic field for a spin system. 
Initially, the system is 
in thermal state $\rho_1 =\sum_n  p_{n}^{(1)}\vert n\rangle \langle n\vert$
at temperature $T_1$, where $p_{n}^{(1)} = \exp(-\beta_1 \varepsilon_n a_1)/Z_1$,
 $\beta_1 = 1/k_{\rm B} T_1$, and the partition function 
 $Z_1 = \sum_n \exp(-\beta_1 \varepsilon_n a_1)$.
The quantum Otto cycle involves the following 
steps \cite{Quan2007}:

(i) The system  is detached
from the hot bath and made to undergo a quantum adiabatic process,
in which the external control is slowly changed from the value $a_1$ to $a_2$.
Thus the hamiltonian changes from $H_1(a_1)$ to $H_2(a_2)$ with eigenvalues
$\varepsilon_{n}^{(2)}= \varepsilon_n a_2$.
Following quantum adiabatic theorem,
the system remains in the instantaneous eigenstate of the hamiltonian
and so the occupation probabilities of the levels remain unchanged. 
For $a_2 < a_1$, this process is the analogue of an adiabatic expansion. 
The work  done {\it by} the system in this stage
is equal to the change in mean energy ${W}_1=  {\rm
Tr}(\rho_1[H_2-H_1])$.
The change in energy spectrum is such that the  
ratio of energy gaps between any two levels before and 
after the quantum adiabatic process is the same. This makes it possible to assign
 temperature to the system  along the adiabatic process. Thus after step (i), this temperature 
is given by $T_1 (a_2/a_1)$.

(ii) The system with changed spectrum $\varepsilon_{n}^{(2)}$ 
is brought to  thermal state 
$\rho(a_2) =\sum_n  p_{n}^{(2)}\vert n\rangle \langle n\vert$
by contact with cold bath at inverse temperature $\beta_2$, where 
$p_{n}^{(2)} = \exp(-\beta_2 \varepsilon_n a_2)/Z_2$ and
$Z_2 = \sum_n \exp(-\beta_2 \varepsilon_n a_2)$.
On average, the heat rejected to the bath
in this step, is defined as $Q_2 = {\rm Tr}([\rho(a_2)-\rho(a_1)] H_2)$.

(iii) The system  is now detached
from the cold bath and made to undergo a second quantum adiabatic process
(compression) during which the control is reset to value $a_1$.
Work done {\it on} the system in this step is  ${W}_2=  {\rm
Tr}(\rho_2[H_1-H_2])$.

(iv) Finally, the system is put in contact with the hot bath again.
 Heat is absorbed by the system in this step, whence it recovers its initial 
 state $\rho_1$. 
On average, the total work done in one cycle, is calculated to be  
\begin{eqnarray}
{W(a_1,a_2)} &=& \sum_{n} \left( \varepsilon_{n}^{(1)}- \varepsilon_{n}^{(2)}
\right) 
\left( p_{n}^{(1)}- p_{n}^{(2)} \right), \\
&=&  (a_1-a_2)\sum_{n} \varepsilon_{n} \left( p_{n}^{(1)}- p_{n}^{(2)} \right) >
 0. 
\label{work}
\end{eqnarray}
Similarly, heat exchanged with hot bath in step (iv) is given by
$Q_1=a_1 \sum_n \varepsilon_n  \left( p_{n}^{(1)}- p_{n}^{(2)} \right) >0.$
Heat exchanged by the system with the cold bath is $Q_2 = {W}-Q_1 <0$.
The efficiency of the engine  $\eta={W }/Q_1$, 
is given by
\begin{equation}
\eta = 1-\frac{a_2}{a_1}.
\label{eta}
\end{equation}
Clearly, this cycle  has two internal energy scales and the efficiency
is also similar to that of Feynman's ratchet, Eq. (\ref{efficiency-ratchet}). One can 
seek an optimal engine configuration, by optimising work output
per cycle over the parameters $a_1$ and $a_2$. However, unlike
the case of Feynman's ratchet as engine, a closed-form expression
for the efficiency at optimal work seems difficult to obtain here \cite{AJM2008}.

We can formulate a problem of estimation here,  for performance
of the engine, assuming that the absolute magnitudes of internal scales
are not known. Further, we simplify by assuming
that the ratio of energy scales, or in other words, the efficiency
is specified. In the following, we briefly outline the emergence
of CA efficiency in this problem. The following treatment 
generalizes the analysis of Ref. \cite{Johal2010}.

It is convenient to express ${W(a_1,a_2)} \equiv {W}(a_1,\eta)$, using Eq.
(\ref{eta}). Due to analogy with the ratchet problem, we may
take the prior for the uncertain parameter $a_1$ to be Jeffreys' prior:
$\Pi(a_1) = N/a_1$, where $N = [\ln(a_{\rm max}/a_{\rm min})]^{-1}$.
 The expected work per cycle 
for a given $\eta$, is then given by
\begin{eqnarray}
\overline{W}(\eta) &=& \int_{a_{\rm min}}^{a_{\rm max}}
{W}(a_1,\eta) \Pi (a_1) d a_1 \\
&=& N \eta  \int_{a_{\rm min}}^{a_{\rm max}} 
\sum_{n} \varepsilon_{n}
\left( p_{n}^{(1)}- p_{n}^{(2)} \right) d a_1
\label{avw} 
\end{eqnarray}
To perform the integration, we write 
$\int \sum_{n} \varepsilon_{n}  p_{n}^{(1)} d a_1 = 
\sum_{n}  \varepsilon_{n} \int p_{n}^{(1)} d a_1$
and integrate by parts. The result can 
be written as: 
\begin{equation}
 \sum_{n}  \varepsilon_{n}\int_{a_{\rm min}}^{a_{\rm max}}  p_{n}^{(1)} d a_1 =
 -\frac{1}{\beta_1} \ln \left(\frac{ \sum_{k=1}^M e^{-\beta_1 \varepsilon_k a_{\rm max}}}
 { \sum_{k=1}^M e^{-\beta_1 \varepsilon_k a_{\rm min}}}  \right). 
\end{equation}
Thus the average work is evaluated to be 
\begin{equation}
 \overline{W} = N \eta\left[\frac{1}{\beta_2(1-\eta)} \ln \left(\frac{ \sum_k e^{-\beta_2(1-\eta)
 \varepsilon_k a_{\rm max}}}
 { \sum_k e^{-\beta_2(1-\eta) \varepsilon_k a_{\rm min}}}  \right)
 -\frac{1}{\beta_1} \ln \left(\frac{ \sum_k e^{-\beta_1 \varepsilon_k a_{\rm max}}}
 { \sum_k e^{-\beta_1 \varepsilon_k a_{\rm min}}}  \right)    \right], 
\label{wev}
 \end{equation}
or, which is written briefly as:
\begin{equation}
  \overline{W} = N \eta\left[\frac{X_2}{\beta_2(1-\eta)} - \frac{X_1}{\beta_1} \right],
\end{equation}
where $X_1$ and $X_2$ can be easily identified from Eq. (\ref{wev}).

Now we wish to find the efficiency at optimal average work, and 
 so we apply the condition 
\begin{equation}
 \frac{\partial \overline{W}}{\partial \eta} = 0 \quad \Longrightarrow \quad 
\frac{X_2}{\beta_2(1-\eta)^2} - \frac{X_1}{\beta_1} + \frac{\eta}{\beta_2 (1-\eta)} 
\frac{\partial X_2}{\partial \eta}= 0.
\label{wzero}
\end{equation}
The resulting equation is, in general, a function of $a_{\rm max}$ and $a_{\rm min}$.
However, we are interested in the asymptotic limit of large $a_{\rm max}$ and vanishing $a_{\rm min}$.
In this limit, 
the dominant term in the sum $\sum_k e^{-\beta_2(1-\eta)
 \varepsilon_k a_{\rm max}}$  is given by $e^{-\beta_2(1-\eta)
 \varepsilon_1 a_{\rm max}}$, where $\varepsilon_1$ is the ground-state energy.
Therefore, 
$X_2 \to -\beta_2 (1-\eta) \varepsilon_1 a_{\rm max} -\ln M$.
Similarly, in the said limit
\begin{equation}
X_1 \to -\beta_1 \varepsilon_1 a_{\rm max} -\ln M, \quad 
 \frac{\partial X_2}{\partial \eta} \to \beta_2 \varepsilon_1 a_{\rm max}.
 \end{equation} 
Finally, using the above limiting forms in Eq. (\ref{wzero}), we obtain:
\begin{equation}
\ln M \left[ \frac{1}{\beta_1} - \frac{1}{\beta_2(1-\eta)^2}   \right] = 0, 
\end{equation}
which implies that the expected work becomes optimal at $\eta = 1-\sqrt{\frac{\beta_1}{\beta_2}}$,
or at CA-efficiency. 
%
\section{Summary}
We observed in Feynman's ratchet that for small temperature differences,
the figures of merit at optimal values of $\overline{P}$ and $\overline{\chi}$,
agree with the corresponding expressions at the optimal
values of $P$ and $\chi$. The important conditions which 
hold in this comparison are,  Jeffreys'
prior as the underlying prior and an asymptotic range of values over which 
the prior is defined. In contrast, the uniform
prior is not able to generate the optimal behavior in
the near equilibrium regime. 
Further we note that for endoreversible models with a Newtonian heat
flow between a reservoir and the working medium,
the efficiency at optimal power is exactly $\eta^*$ \cite{CA1975, Joubook}.
Correspondingly, the COP at optimal $\chi$-criterion    
is given by $\zeta^*$ \cite{Chen1990}. In this paper, these values are obtained
with an inference based approach assuming incomplete information
in a mesoscopic model of heat engine. We have also shown that
our analysis applies to a broader class of idealized models of heat engines/refrigerators,
driven by quasi-static processes. Here also, CA efficiency 
emerges from the use of Jeffreys' prior, under the given conditions of the model. 
 
We conclude with  an argument to support as to why our approach yields the familiar results
of finite-time thermodynamics. 
To exemplify, in the case of Feynman's ratchet, 
the asymptotic range has been considered {\it after} we
optimized the expected power output (in case of engine) over 
the efficiency. One may consider these two 
steps in the opposite order, i.e. take the asymptotic range
first and then perform the optimization. For that
we rewrite Eq. (\ref{q1dot}) as follows:
\begin{equation}
\dot{Q}_1 (\epsilon_2,\eta) =\frac{r_0 \epsilon_2}{(1-\eta)}
\left(e^{-\epsilon_2/(1-\eta)T_1}- e^{-\epsilon_2/T_2}\right),
\end{equation}
and define the expected heat flux as
\begin{equation}
\overline{\dot{Q}}_1 = \int_{\epsilon_{\rm min}}^{\epsilon_{\rm max}}
\dot{Q}_1 (\epsilon_2,\eta) \Pi(\epsilon_2) d\epsilon_2.
\end{equation}
Then in the asymptotic range, we obtain the approximate expression as
\begin{equation}
 \overline{\dot{Q}}_1 \approx C\left[ T_1 - \frac{T_2}{1-\eta} \right],
\end{equation}
where $C$ is as in Eq. (\ref{defc}). Here we can   
draw a parallel with Newtonian heat flow: $\overline{\dot{Q}}_1 \propto [T_1-T_1']$
where $T_1' \equiv {T_2}/{(1-\eta)}<T_1$ is an effective temperature. 
Similarly, the prior-averaged rate of heat rejected
to the cold reservoir can be written as
\begin{equation}
 \overline{\dot{Q}}_2 \approx C\left[ (1-\eta){T_1} - T_2 \right].
\end{equation}
Here also, we may identify another Newtonian heat flow  
$\overline{\dot{Q}}_2 \propto [T_2'-T_2]$, with the same effective heat conductance $C$,
between an effective temperature 
$T_2'\equiv (1-\eta){T_1}>T_2$ and temperature $T_2$ of the cold reservoir.
Then it is easily seen that the maximum of expected power  
$\overline{P} = \overline{\dot{Q}}_1 - \overline{\dot{Q}}_2$,
is obtained at CA value.
Similarly, one can argue for the emergence of $\zeta^*$
in the case of refrigerator mode, in terms of effective heat flows
which are Newtonian in nature. 

Interestingly, the above expressions 
seem to suggest an analogy between the expected
mesoscopic model with limited information,
and a finite-time thermodynamic model  
with Newtonian heat flows.
If we compare with the endoreversible models \cite{CA1975, Chen1990},
then we observe that the assumption of a Newtonian heat flow
goes together with obtaining CA efficiency
at maximum power, and COP $\zeta^*$ 
at optimum $\chi$-criterion. We however note 
that the analogy does not hold in entirety.
The effective temperatures defined above do not
have physical counterpart in the ratchet model,
although in the endoreversible picture, these
denote the temperatures of the working medium
while in contact with hot or cold reservoirs. 
Secondly, the heat conductances need not be equal 
for the endoreversible model with Newtonian heat flows.
Further, the intermediate temperatures
$T_1'$ and $T_2'$ as above, are equal in magnitude at the maximum expected power $\overline{P}$.
However, for the endoreversible model, these temperatures
are not equal at maximum power \cite{CA1975, Joubook}.
Still, the form of expressions for the rates of heat transfer do provide 
a certain insight into the emergence of the familiar expressions
for figures of merit at optimal expected performance within the 
prior-averaged approach. 

Finally, we close with a few observations on future lines
of enquiry.
It was seen in Fig. 1, that for a specified finite range for the prior, the estimates of
efficiency at maximum power are either above, or below the estimates in the asymptotic 
range. In particular,  the estimates
are function of the values $\epsilon_{\rm min}$ and $\epsilon_{\rm max}$.
We obtain universal results, dependent on the ratio of  reservoir
temperatures, only in the asymptotic range.
Further, the smaller values of the upper limit, overestimate
the efficiency (inset in Fig. 1) whereas the larger values of the lower limit,
underestimate the efficiency. An opposite behavior is seen
for the refrigerator mode (Fig. 4). Moreover, this trend for a 
chosen mode (engine/refrigerator) is specific to
the choice of the uncertain variable.
Thus the trend is reversed, if instead of choosing $\epsilon_2$, 
we perform the analysis with $\epsilon_1$ as the uncertain variable. 
This behavior is seen in both the engine as well as the refrigerator mode.
Investigation into the relation between inferences derived from 
the two choices for the uncertain variable, may yield further insight
into the behavior of estimated performance and the approach in general.
The point may be appreciated by noting that by specifying a finite-range for the prior
we add new information to the probabilistic model. In order that inference
may provide a useful and practical guess on the actual performance
of the device, this additional prior information has to be
related to some objective features of the model. These considerations
are relevant for further exploring the intriguing relation between the subjective
and the objective descriptions of thermodynamic models \cite{JRM2014}. 
\section{Acknowledgement}
The authors acknowledge financial support from 
the Department of Science and Technology,
India under the research project No. SR/S2/CMP-0047/2010(G), titled:
``Quantum Heat Engines: work, entropy and information at the nanoscale''.


\begin{thebibliography}{10}
\expandafter\ifx\csname url\endcsname\relax
  \def\url#1{\texttt{#1}}\fi
\expandafter\ifx\csname urlprefix\endcsname\relax\def\urlprefix{URL }\fi
\expandafter\ifx\csname href\endcsname\relax
  \def\href#1#2{#2} \def\path#1{#1}\fi

\bibitem{CA1975}
F.~L. Curzon, B.~Ahlborn, Efficiency of a {C}arnot engine at maximum power
  output, Am. J. Phys. 43 (1975) 22.

\bibitem{DeVos1992}
A.~De~Vos, Endoreversible Thermodynamics of Solar Energy Conversion, Oxford
  science publications, Oxford University Press, Oxford, 1992.


\bibitem{Salamon2001}
P.~Salamon, J.~Nulton, G.~Siragusa, T.~Andersen, A.~Limon,
  {Principles of control thermodynamics}, Energy 26~(3) (2001) 307.
  
 
  \bibitem{Broeck2010}
M.~Esposito, R.~Kawai, K.~Lindenberg, C.~Van~den Broeck,
  {Efficiency at
  maximum power of low-dissipation {C}arnot engines}, Phys. Rev. Lett. 105
  (2010) 150603.
  

  
\bibitem{Broeck2005}
C.~Van~den Broeck,
  {Thermodynamic efficiency at maximum power}, Phys. Rev. Lett. 95 (2005) 190602.


\bibitem{Broeck2009}
M.~Esposito, K.~Lindenberg, C.~Van~den Broeck,
  {Universality of
  efficiency at maximum power}, Phys. Rev. Lett. 102 (2009) 130602.


\bibitem{Chen1990}
Z.~Yan, J.~Chen, {A class of
  irreversible {C}arnot refrigeration cycles with a general heat transfer law},
  J. Phys. D: Appl. Phys. 23~(2) (1990) 136.

  
  \bibitem{Apertet}
Y.~Apertet, H.~Ouerdane, A.~Michot, C.~Goupil, P.~Lecoeur,
  {On the efficiency at
  maximum cooling power}, Europhys. Lett. 103~(4) (2013) 40001.


\bibitem{Armen2010}
A.~E. Allahverdyan, K.~Hovhannisyan, G.~Mahler,
  {Optimal
  refrigerator}, Phys. Rev. E 81 (2010) 051129.


\bibitem{Tomas2012}
C.~de~Tom\'as, A.~C. Hern\'andez, J.~M.~M. Roco,
  {Optimal low
  symmetric dissipation {C}arnot engines and refrigerators}, Phys. Rev. E 85
  (2012) 010104.


\bibitem{Roco2012}
Y.~Wang, M.~Li, Z.~C. Tu, A.~C. Hern\'andez, J.~M.~M. Roco,
  {Coefficient of
  performance at maximum figure of merit and its bounds for low-dissipation
  carnot-like refrigerators}, Phys. Rev. E 86 (2012) 011127.

\bibitem{Roco2013}
Y.~Hu, F.~Wu, Y.~Ma, J.~He, J.~Wang, A.~C. Hern\'andez, J.~M.~M. Roco,
  {Coefficient of
  performance for a low-dissipation {C}arnot-like refrigerator with nonadiabatic
  dissipation}, Phys. Rev. E 88 (2013) 062115.
  
\bibitem{Jeffreys1939}
H.~Jeffreys, Theory of Probability, Clarendon Press, Oxford, 1939.  
  
  
 \bibitem{Jaynes1968}
E.~Jaynes, Prior probabilities, IEEE Trans. Syst. Sci. Cybernet. 4 (1968) 227.

\bibitem{Johal2010}
R.~S. Johal,
  {Universal
  efficiency at optimal work with {B}ayesian statistics}, Phys. Rev. E 82
  (2010) 061113.


\bibitem{GJ2012}
G.~Thomas, R.~S. Johal,
  {Expected behavior
  of quantum thermodynamic machines with prior information}, Phys. Rev. E 85
  (2012) 041146.


\bibitem{GPJ2012}
G.~Thomas, P.~Aneja, R.~S. Johal,
  {Informative
  priors and the analogy between quantum and classical heat engines}, Physica
  Scripta (T151) (2012) 014031.


\bibitem{PJ2013}
P.~Aneja, R.~S. Johal,
  {Prior information and
  inference of optimality in thermodynamic processes}, J. Phys. A:
  Math. Theor. 46~(36) (2013) 365002.

  \bibitem{RJ2014}  R.~S. Johal, 
  Efficiency at optimal work from finite source and sink: a probabilistic perspective, 
	 J. Noneq. Therm. 40 (2015) 1.

   \bibitem{Seifert}
T.~Schmiedl, U.~Seifert,
  {Efficiency at maximum
  power: An analytically solvable model for stochastic heat engines}, 
  Europhys. Lett. 81~(2) (2008) 20003.
  
  \bibitem{Zhang2006} Y. Zhang, B.~H. Lin, J.~C. Chen, 
Performance characteristics of an irreversible thermally driven Brownian microscopic heat engine.
Eur. Phys. J. B 53 (2006) 481.

\bibitem{BaratoSeifert} A.~C. Barato, U. Seifert, An autonomous and reversible Maxwell's demon.
Europhys. Lett. 101 (2013) 60001.
  
  \bibitem{JRM2014} R.~S. Johal, R. Rai, G. Mahler,
  Reversible heat engines: Bounds on estimated efficiency from inference, 	     
	   Found. Phys. 45 (2015) 158.

\bibitem{Feynman1966}
R.~P. Feynman, R.~B. Leighton, M.~Sands, The Feynman Lectures on Physics,
  Addison-Wesley, Reading, MA, 1966.
  
 \bibitem{Hernandez}
S.~Velasco, J.~M.~M. Roco, A.~Medina, A.~C. Hernández,
 {Feynman's ratchet
  optimization: maximum power and maximum efficiency regimes}, J.
  Phys. D: Appl. Phys. 34~(6) (2001) 1000.


\bibitem{Tu2008}
Z.~C. Tu, {Efficiency at
  maximum power of Feynman's ratchet as a heat engine}, J.  Phys. A:
  Math. Theor. 41~(31) (2008) 312003.

  \bibitem{Abe2014} S. Abe,
  Conditional maximum-entropy method for selecting prior
distributions in Bayesian statistics, Europhys. Lett. 108 (2014) 40008.
  
    
\bibitem{He2013}
X.~G. Luo, N.~Liu, J.~Z. He,
  {Optimum analysis of
  a {B}rownian refrigerator}, Phys. Rev. E 87 (2013) 022139.


\bibitem{Tu2014}
S.~{Sheng}, P.~{Yang}, Z.~C. {Tu}, {Coefficient of performance at maximum
  $\chi$-criterion for {F}eynman ratchet as a refrigerator},
  Commun. Theor. Phys. 62 (2014) 589. 
 
\bibitem{Liu2006}
B.-Q. Ai, L.~Wang, L.-G. Liu,
  {Brownian micro-engines and refrigerators in a spatially periodic temperature
  field: Heat flow and performances}, Phys. Lett. A 352~(4–5) (2006)
  286.


\bibitem{Chen2009}
B.~Lin, J.~Chen,
  {Performance
  characteristics and parametric optimum criteria of a brownian
  micro-refrigerator in a spatially periodic temperature field}, 
  J. Phys. A: Math. Theor. 42~(7) (2009) 075006.


\bibitem{Buttiker1987}
M.~Büttiker, {Transport as a
  consequence of state-dependent diffusion}, Zeitschrift für Physik B
  Condensed Matter 68~(2-3) (1987) 161.


\bibitem{Landauer1988}
R.~Landauer, {Motion out of noisy
  states}, J.  Stat. Phys. 53~(1-2) (1988) 233.
  
 \bibitem{ShengTu}
S.~Sheng, Z.~C. Tu,
 {Universality of
  energy conversion efficiency for optimal tight-coupling heat engines and
  refrigerators}, J. Phys. A: Math. Theor. 46~(40)
  (2013) 402001.
  
  \bibitem{Kieu2004} T.D. Kieu, Quantum heat engines, the second law and Maxwell's daemon,
  Eur. Phys. J. D 39 (2006) 115.
  
  \bibitem{Quan2007} H.T. Quan,Yu-xi Liu, C. P. Sun, and Franco Nori,
  Quantum thermodynamic cycles and quantum heat engines, 
  Phys. Rev. E 76 (2007) 031105.
  
  \bibitem{AJM2008} A. Allahverdyan, R.S. Johal, G. Mahler, 
  Work extremum principle: Structure and function of quantum heat engines,
  Phys. Rev. E 77 (2008) 041118.
  
  \bibitem{Joubook} G. Lebon, D. Jou, J. Casas-Vásquez, Understanding Nonequilibrium
  Thermodynamics, Springer, Berlin, 2008.
\end{thebibliography}
\end{document}